\documentclass[journal]{IEEEtran}

\usepackage[breaklinks=true]{hyperref}
\usepackage{breakcites}
\usepackage{amssymb}
\usepackage{amsmath}
\usepackage{graphicx} 
\usepackage{xcolor}
\usepackage{subcaption}
\usepackage{booktabs}

\usepackage{algorithm}
\usepackage{algpseudocode}

\newtheorem{assumption}{Assumption}
\newtheorem{attack}{Attack scenario}
\ifCLASSINFOpdf
\else
\fi
\usepackage[final]{changes}
\usepackage{amsfonts}

\begin{document}
%
\title{\replaced{SwarmRaft}{SwarnRaft}: Leveraging Consensus for Robust Drone Swarm Coordination in GNSS-Degraded Environments}
%
%
%


\author{
Kapel Dev,
Yash Madhwal,~\IEEEmembership{Member,~IEEE}
Sofia Shevelo, 
Pavel Osinenko, 
Yury Yanovich,~\IEEEmembership{Member,~IEEE}
\thanks{All the authors are from Skolkovo Institute of Science and Technology.}
\thanks{Manuscript received xx, 2025; revised xx, 2025.}
}

%
%

\markboth{IEEE INTERNET OF THINGS JOURNAL,~Vol.~NN, No.~N, August~2025}%
{Shell \MakeLowercase{\textit{et al.}}: Bare Demo of IEEEtran.cls for IEEE Journals}
%



\maketitle



\begin{abstract}
Unmanned aerial vehicle (UAV) swarms are increasingly used in critical applications such as aerial mapping, environmental monitoring, and autonomous delivery. However, the reliability of these systems is highly dependent on uninterrupted access to the Global Navigation Satellite Systems (GNSS) signals, which can be disrupted in real-world scenarios due to interference, environmental conditions, or adversarial attacks, causing disorientation, collision risks, and mission failure. This paper proposes \textbf{\replaced{SwarmRaft}{SwarnRaft}}, a blockchain-inspired positioning and consensus framework for maintaining coordination and data integrity in UAV swarms operating under GNSS-denied conditions. \replaced{SwarmRaft}{SwarnRaft} leverages the Raft consensus algorithm to enable distributed drones (nodes) to agree on state updates such as location and heading, even in the absence of GNSS signals for one or more nodes. In our prototype, each node uses GNSS and local sensing, and communicates over WiFi in a simulated swarm. Upon signal loss, consensus is used to reconstruct or verify the position of the failed node based on its last known state and trajectory. Our system demonstrates robustness in maintaining swarm coherence and fault tolerance through a lightweight, scalable communication model. This work offers a practical and secure foundation for decentralized drone operation in unpredictable environments.
\end{abstract}

\begin{IEEEkeywords}
Blockchain, UAV swarms, Raft consensus, GNSS-denied environments, Fault tolerance
\end{IEEEkeywords}

%
\IEEEpeerreviewmaketitle

\section{Introduction}
Rapid proliferation of autonomous swarms, including unmanned aerial vehicles (UAVs), ground robots, and maritime systems, has enabled transformative applications in domains such as logistics, infrastructure monitoring, agriculture, and disaster response \cite{Laghari2023,AlMarshoud2024,Bae2023}. These distributed systems rely heavily on precise coordination, predominantly facilitated by \textit{Global Navigation Satellite Systems} (GNSS) such as \textbf{GPS}, \textbf{GLONASS}, \textbf{Galileo}, and \textbf{BeiDou} \cite{Bhatta2021}. However, GNSS signals are often unreliable in real-world conditions, such as urban canyons, dense vegetation, or indoor and subterranean environments. \added{Even with recent advances in noise-resilient GNSS acquisition techniques \cite{Zhang2023}, the technology remains vulnerable to spoofing, jamming, and environmental interference.} To mitigate these limitations, systems frequently integrate \textit{Inertial Navigation Systems} (INS) to improve robustness and continuity. \added{This inherent limitation of standalone GNSS or INS systems motivates the need for collaborative, consensus-driven approaches, such as SwarmRaft, that combine GNSS, INS, and peer-to-peer fusion to achieve robust and fault-tolerant state estimation in UAV swarms.}

As swarm deployments scale in size and mission complexity, achieving reliable consensus among agents becomes a critical challenge, one that draws direct parallels to problems in distributed computing and blockchain technologies. In this context, swarms must maintain accurate and synchronized state information (e.g., position, velocity, heading) despite partial failures and environmental interference \cite{Hoang2023}\deleted{,\cite{Elfatih2023}}.

Several mission-critical scenarios exemplify the need for robust consensus mechanisms. In urban package delivery, GNSS reflections from buildings can compromise altitude coordination, and a single altimeter failure can lead to collisions~\cite{Kuru2019}. During bridge inspections, complete GNSS denial requires drones to maintain spatial awareness solely through local sensing and swarm coordination. Agricultural spray missions present additional complexities, as drones must adjust for dynamic weight changes during chemical dispersion in rural settings, often degraded by GNSS.

These operational constraints echo classic challenges of distributed systems, yet swarm robotics introduces distinct requirements~\cite{Bano2019,Raikwar2024}. Unlike traditional consensus algorithms that emphasize Byzantine fault tolerance, such as those used in permissioned or permissionless blockchain systems, swarm consensus must prioritize low-latency agreement, operate under stringent resource constraints, and address crash-fault-dominated failure models~\cite{Buterin2015,Bitfury2015,Bitfury2015a}. The core principles of consensus remain: ensuring \textit{termination} (all non-faulty agents reach a decision), \textit{agreement} (consistency across decisions), and \textit{integrity} (decisions are based on valid inputs), even in the presence of partial system failures \cite{Dwork1988}.

UAV swarms operate under unique constraints, including limited compute and energy resources, intermittent connectivity, and bounded-delay communication networks. Consequently, traditional blockchain consensus protocols such as Proof-of-Work and Proof-of-Stake are ill-suited due to their computational and energy overhead~\cite{Nakamoto2008,Kiayias2017}. Similarly, Byzantine Fault-Tolerant (BFT) protocols, e.g., \replaced{Practical Byzantine Fault Tolerance (PBFT)}{PBFT}, Tendermint, Exonum \cite{Castro1999,Kwon2014,Exonum2018}, assume adversarial behavior and incur unnecessary complexity for typical crash-fault scenarios in swarms. This creates a compelling opportunity to adapt crash-tolerant consensus protocols, such as \textit{Raft} \cite{Ongaro2014}, which are designed for environments where nodes (the drones) may fail by crashing, but not by acting maliciously. Raft's lightweight leader-based architecture is well suited to the resource-constrained and real-time nature of swarm robotics.

To address these challenges, we propose \textbf{SwarmRaft}, a \textit{consensus-driven} positioning and \textit{crash-tolerant} system designed for drone swarms. Our approach leverages the Raft consensus algorithm to enable drones to communicate and synchronize over a distributed network, even in partially GNSS-denied environments. SwarmRaft integrates GNSS and INS data to enable drones to exchange critical state information such as position and heading. In the event of GNSS loss or sensor malfunction, the swarm uses consensus to reconstruct or verify the location and trajectory of affected nodes based on shared data and prior motion. This consensus-driven estimation ensures that the swarm remains cohesive and continues its mission safely, even when individual drones experience degraded sensing.

Our work makes the following key contributions:  
\begin{itemize}  
    \item \textbf{Adaptation of Raft for Swarm Environments}: We adapt the Raft consensus protocol to support real-time coordination in UAV swarms, prioritizing low-latency decision-making and resilience under crash-fault conditions typical in GNSS-compromised settings.
    \item \textbf{Sensor Fusion for Robust State Estimation}: SwarmRaft combines GNSS and INS data using a distributed consensus framework to produce robust, fault-tolerant state estimates. This fusion mitigates the effects of signal intermittency, spoofing, and sensor drift, enhancing the swarm's overall situational awareness and operational reliability.
    \item \textbf{Empirical Validation in Realistic Scenarios}: We evaluate SwarmRaft through a comprehensive set of experiments and simulations, demonstrating its effectiveness in preserving swarm cohesion and operational continuity across urban and natural terrains.
\end{itemize}

\added{Motivated by the real-world challenges UAV swarms face, such as GNSS spoofing, jamming, and sensor failure, we develop a lightweight, fault-tolerant coordination framework that maintains collective situational awareness even when individual agents are compromised. Unlike centralized or resource-intensive approaches, \textbf{SwarmRaft} leverages crash-tolerant consensus and sensor fusion to operate efficiently in contested or degraded environments. This work bridges swarm robotics and distributed consensus, demonstrating that established protocols like Raft can be adapted for real-time UAV coordination. Our design balances theoretical soundness with practical deployability, yielding substantial improvements in localization accuracy and swarm resilience across simulated and real-world scenarios. SwarmRaft demonstrates how lightweight consensus protocols can enhance resilience and coordination in UAV swarms operating under GNSS-degraded or adversarial conditions.}

\deleted{This work bridges the fields of swarm robotics and distributed consensus, advancing resilience in GNSS-degraded environments. SwarmRaft builds on established consensus algorithms, demonstrating their applicability to UAV swarm coordination. Our approach maintains theoretical rigor while prioritizing practical deployability, showing significant improvements in both simulated and real-world scenarios.}

\section{Related Work}

Decentralized coordination mechanisms for UAV swarms have gained increasing attention, particularly through consensus-based and blockchain-backed approaches. While leader-follower architectures have shown success in formation control~\cite{rafifandi2019leader}, their reliance on central nodes introduces vulnerabilities to single points of failure. In contrast, distributed strategies offer greater resilience. Tariverdi et al.~\cite{tariverdi2023raftingconsensusformationcontrol} present a formation control framework tailored to UAV dynamics, and Jia et al.~\cite{jia2021distributed} demonstrate decentralized estimation and coordination in quadcopter swarms. \added{Similarly, Zuo et al.~\cite{Zuo2022} propose a voting-based scheme for leader election in lead–follow UAV swarms under constrained communication ranges, representing an earlier Raft-inspired approach tailored to leader selection rather than distributed fault-tolerant localization.}

In the context of Industrial IoT \added{(IIoT)}, Seo et al.~\cite{Seo2021} propose a co-design framework that jointly optimizes communication scheduling and consensus execution to meet strict latency requirements. Although their setting involves fixed infrastructure, their insights into delay-aware consensus coordination are applicable to mobile multi-agent systems. Similarly, Ilić et al.~\cite{Ilić2025} introduce adaptive asynchronous gossip algorithms for heterogeneous sensor networks, offering resilience and scalability through weighted message exchange without requiring global synchronization. While their work focuses on distributed estimation in sensor-rich environments, the underlying principles of asynchronous fault-tolerant consensus motivate swarm-scale extensions.

Blockchain-based protocols add guarantees of trust and tamper resistance in decentralized coordination. P-Raft~\cite{lu2023p} optimizes consensus in consortium settings, and \replaced{Dynamic Trust Practical Byzantine Fault Tolerance (DTPBFT)}{DTPBFT}~\cite{han2024dtpbft} dynamically adjusts trust levels in UAV networks. Yazdinejad et al.~\cite{Yazdinejad2021} propose a zone-based drone authentication framework using a Delegated Proof-of-Stake (DDPOS) mechanism to enable secure, low-latency inter-zone handovers in smart city deployments\added{, and fairness-oriented multi-UAV communication frameworks for urgent tasks~\cite{Xu2025}}.

Comparative analyses of consensus algorithms\added{,} such as Raft, PBFT, and \replaced{Proof-of-Work (PoW)}{PoW} have focused on trade-offs in scalability, fault tolerance, and energy use~\cite{sharma2020novel,Androulaki2018,Kostyuk2020,Borzdov2023} \added{, and communication efficiency optimizations in UAV ad hoc networks~\cite{Chen2024}}. Surveys such as~\cite{aditya2021survey} highlight the challenges of adapting these mechanisms to real-time robotics, particularly under energy and processing constraints.

Research on fault-tolerant swarm behavior includes predictive failure mitigation~\cite{o2023predictive}, self-healing topologies~\cite{varadharajan2020swarm}, and GNSS spoofing resilience~\cite{ranganathanimpact}\added{, as well as UAV-aided localization via RF characteristics~\cite{10644103}}. Robust navigation under sensor failures~\cite{hu2022fault} and adaptive topology control~\cite{wang2017fault} further emphasize the need for distributed, resilient solutions.

Real-world deployments, such as outdoor flocking and coordinated flight demonstrations~\cite{vasarhelyi2014outdoor}, underline the feasibility of translating algorithmic designs to physical swarm systems.

Despite these advances, many existing frameworks either impose significant communication or computational overhead, or fail to address the specific fault models and coordination challenges faced by UAV swarms operating under GNSS degradation and adversarial interference. In contrast, \replaced{SwarmRaft}{\textbf{SwarmRaft}} introduces a lightweight, crash-tolerant localization framework that combines peer-based voting with distributed fault recovery, offering robust performance in dynamic and partially compromised environments.


\section{Proposed Solution}

\subsection{Overview of the Consensus Scheme}
In modern UAV swarms, individual GNSS or INS readings alone cannot guarantee resilience against sensor faults or deliberate spoofing. Our consensus scheme, \replaced{SwarmRaft}{\textbf{SwarmRaft}}, leverages peer-to-peer distance measurements, crash fault-tolerant communication consensus, and a Byzantine-resilient evaluation mechanism to detect and correct malicious or \textit{faulty} position reports. At each time step \( k \), the swarm determines the position of each UAV (node). To determine the positions, nodes communicate via Raft consensus and elect a leader for the step. \replaced{The leader collects every node’s position estimate (derived from GNSS when available and maintained by INS otherwise), tests if it is consistent with others, recovers it if required, and returns the results to the nodes.}{The leader collects every node's raw GNSS and INS fused position, tests if it is consistent with others, recovers it if required, and returns the results to the nodes.}  

This design achieves fault tolerance under the assumption that up to \( f \) nodes' sensors may be corrupted, so long as \( n \geq 2f + 1 \), where \( n \) is the total number of nodes in the swarm. \textit{Honest} nodes never collude to mislead, and initial positions are trusted. The result is a distributed protocol that detects spoofing and significant sensor drift, locally filters out unreliable data, and recovers accurate positions through consensus based on neighbor estimates.

A high-level workflow of \replaced{SwarmRaft}{\textbf{SwarmRaft}} in the case of a determined leader at a single step \(k\) is as follows: 
\begin{enumerate}
  \item \textbf{Sense}: Each UAV measures \replaced{its position(GNSS/INS) and internode distances}{GNSS and INS}.
  \item \textbf{Inform}: \replaced{Measurement data}{Raw fused position} is sent to the leader (as a client's transaction).
  \item \textbf{Estimate}: The leader recomputes each UAV's position from range constraints.
  \item \textbf{Evaluate}: The leader determines if the data of each sensor is \textit{faulty} or \textit{honest} based on residuals.
  \item \textbf{Recover}: If a sensor is determined to be \textit{faulty}, the leader replaces its data with the \replaced{position computed with}{median of} peer estimates.
  \item \textbf{Finalize}: The leader sends the final results to all nodes.
\end{enumerate}

\subsection{System \& Threat Model}

Each UAV in our scheme is modeled as a fully-capable sensing and communication node. At each discrete time step \( k \), node \( i \) collects an absolute position measurement \( \mathbf{z}^{\rm GNSS}_{i,k} \) from its onboard GNSS receiver; because GNSS alone can drift or be spoofed, the UAV also carries an INS that produces short-term motion increments \( \Delta \mathbf{u}_{i,k} \) through accelerometers and gyroscopes. To bind these two modalities and detect inconsistent readings, every UAV is further equipped with a ranging sensor\replaced{, }{(}for example, ultra-wideband or \added{Received Signal Strength Indicator (RSSI)}-based\replaced{)}{,} that yields noisy inter-node distances \( d_{ij,k} \) to each neighbor \( j \). Finally, nodes share all measurements and voting messages over an authenticated broadcast channel protected by pre-distributed cryptographic keys, so that neither message forgery nor tampering can occur undetected.

We assume the communication graph among the \( n \) UAVs is fully connected and synchronous: in each protocol round every node can exchange messages with every other node, and timing is sufficiently well aligned that messages from step \( k \) are received before step \( k+1 \) begins. This idealization lets us focus on sensor faults rather than network delays or partitions. To tolerate up to \( f \) arbitrarily \textit{faulty} or malicious sensors, we require the bound \( n \geq 2f + 1 \). Under this constraint, even if \( f \) nodes report completely corrupted GNSS or ranging data (or collude to bias their votes), the remaining \textit{honest} nodes still form a majority that can detect and override any bad information.

\begin{assumption}
Nodes' compute and communication modules are \textit{honest}, authorized, and synchronous.
\end{assumption}

\begin{assumption}
Only GNSS sensors can be Byzantine, while INS are reliable. Up to \( f \) nodes' GNSS sensors out of \( n \) may be arbitrarily corrupted, where \( n \geq 2f + 1 \).
\end{assumption}

\begin{assumption}
True positions at time \( k=0 \) are securely known.
\end{assumption}

We model three classes of adversarial behavior against which our consensus scheme must defend:

\begin{attack}[GNSS Spoofing Attack]
An adversary corrupts the satellite positioning signals received by up to \( f \) UAVs, causing each compromised node to report position measurements \( \mathbf{z}^{\rm GNSS}_{i,k} \) that are offset by an arbitrary, potentially time-varying bias. Such spoofing can be orchestrated remotely and may remain undetected by the individual UAV's receiver.
\end{attack}

However, our inter-node consistency checks and consensus mechanism aim to reveal any GNSS readings that deviate significantly from the peer-estimated location.

\begin{attack}[Ranging-Tampering Scenario]  
The adversary injects errors into the inter-UAV distance measurements \( d_{ij,k} \). This could be accomplished by jamming or forging ultra-wideband pulses (or manipulating RSSI readings), causing one or more \textit{honest} nodes to compute incorrect range constraints. 
\end{attack}

Since our fused position estimates rely on the integrity of these distance measurements, the protocol's voting and median-based recovery step are designed to mitigate the impact of up to \( f \) such corrupted range reports.

\begin{attack}[Collusion Among Faulty Sensors]  
We allow for collusion among up to \( f \) \textit{faulty} sensors across different UAVs. In this worst-case scenario, compromised nodes coordinate both their reported measurements and their votes in the \replaced{SwarmRaft}{\textbf{SwarmRaft}} protocol in an attempt to sway the swarm's decision about a particular node's status. 
\end{attack}

By enforcing the requirement \( n \geq 2f + 1 \) and using majority thresholds, the scheme guarantees that \textit{honest} votes outnumber colluding malicious votes, ensuring that no coalition of size \( \leq f \) can force an incorrect global decision.

\subsection{Position Estimation}

\begin{table}[ht]
\centering
\caption{\added{Summary of Notation}}
\label{tab:notation}
\begin{tabular}{ll}
\toprule
\textbf{Symbol} & \textbf{Description} \\
\midrule
$x_{i,k}$             & True position of drone $i$ at time $k$ \\
$z_{i,k}^{GNSS}$      & GNSS position measurement of drone $i$ \\
$v_{i,k}^{GNSS}$      & GNSS measurement noise \\
$x_{i,k}^{INS}$       & Position estimated from INS \\
$v_{i,k}^{INS}$       & INS sensor noise or drift \\
$d_{ij,k}$            & Measured distance from drone $i$ to $j$ \\
$\eta_{ij,k}$         & Noise in range measurement $d_{ij,k}$ \\
$R_{GNSS}, R_{INS}$   & Covariance matrices for GNSS and INS noise \\
$\sigma_d^2$          & Variance of range noise \\
$N$                   & Number of drones in the swarm \\
$f$                   & Number of compromised/faulty drones \\
\bottomrule
\end{tabular}
\end{table}

At each time step \( k \in \mathbf{N} \), the UAV with index \( i \in \{1, \dots, n\} \) obtains two complementary \added{sources of information about}{measurements of} its state. First, the GNSS receiver produces a direct but noisy readout of the true position \deleted{\( \mathbf{x}_{i,k} \in \mathbb{R}^3  \)}{\( \mathbf{x}_{i,k}\)}:  
\[
  \mathbf{z}^{\rm GNSS}_{i,k} \;=\;\mathbf{x}_{i,k} \;+\;\mathbf{v}^{\rm GNSS}_{i,k},
  \quad \mathbf{v}^{\rm GNSS}_{i,k}\sim\mathcal{N}(0,R_{\rm GNSS}),
\]
\replaced{where $z_{i,k}^{\text{GNSS}}$ is the GNSS measurement, 
$v_{i,k}^{\text{GNSS}} \in \mathbb{R}^3$ is the GNSS measurement noise, and 
$R^{\text{GNSS}} \in \mathbb{R}^{3 \times 3}$ is its covariance matrix. 
Here, $\mathcal{N}(\mu, \Sigma)$ denotes a multivariate Gaussian distribution 
with mean $\mu$ and covariance $\Sigma$.}{where \( R_{\rm GNSS} \) is the GNSS measurement covariance. and \(\mathcal{N}(m,\sigma^2)\) is a normal distribution with mathematical expectation $m$ and variance $\sigma^2$.}

\replaced{Second, in the event of a GNSS outage, the UAV can fall back on its inertial navigation system. 
The INS predicts the next position \( \mathbf{x}_{i,k}^{INS} \in \mathbb{R}^3  \)  based on the previous estimate 
$x_{i,k-1}$ and the inertial measurements $\Delta u_{i,k}$, 
which represent increments derived from accelerometer and gyroscope readings 
over the interval $[k-1, k]$:
}
{Second, the INS dead-reckons from the previous time step by integrating acceleration and rotation increments \( \Delta\mathbf{u}_{i,k} \), yielding}

\[
\added{x_{i,k}^{\text{INS}} = f\!\left(x_{i,k-1}, \, \Delta u_{i,k}\right) + v_{i,k}^{\text{INS}}, 
\quad v_{i,k}^{\text{INS}} \sim \mathcal{N}(0, R^{\text{INS}}),}
\] 

\added{where $f(\cdot)$ denotes the INS state-propagation model, 
$v_{i,k}^{\text{INS}} \in \mathbb{R}^3$ is the accumulated INS error over one step, and $R^{\text{INS}} \in \mathbb{R}^{3 \times 3}$ characterizes the associated covariance.}

Thus, the INS provides high-rate relative motion at the cost of accumulating drift, while GNSS provides drift-free absolute position with higher instantaneous noise.

To cross-check these two sensor streams, each UAV also measures its distance to every other UAV. If UAV \( j \) measures the range to UAV \( i \), the reading satisfies  
\[
  d_{ij,k}
  = \bigl\|\mathbf{x}_{i,k}-\mathbf{x}_{j,k}\bigr\|
    + \eta_{ij,k},
  \quad \eta_{ij,k}\sim\mathcal{N}(0,\sigma_d^2).
\]
\added{where $d_{ij,k}$ is the measured distance from UAV $j$ to UAV $i$ at time step $k$, 
$x_{i,k}$ and $x_{j,k}$ are the true positions of UAVs $i$ and $j$, respectively, 
$\eta_{ij,k}$ is the measurement noise, and $\sigma_d^2$ is its variance.}This geometric constraint links the true positions of nodes \( i \) and \( j \) and can reveal inconsistencies when one node's GNSS \replaced{add/or}{or} INS has been spoofed or drifted excessively.

\subsection{\replaced{Fused Estimation through Probabilistic Localization}{Fused Estimate at Peer \( j \) for Node \( i \)}}

\added{Each node \( i \) measures its distances to all neighboring nodes and pairs this information with its own GNSS-derived position. Both the GNSS estimate and the set of computed inter-node distances are then transmitted to the leader node. Similarly, every other node in the network reports its GNSS position together with its locally computed distances to neighbors.
Upon receiving the report from node \( i \), the leader uses the GNSS location of \( i \) and its distance measurements to approximate the possible locations of neighboring nodes. Since a single distance measurement corresponds to a circle (in 2D) or a sphere (in 3D) centered at node \( i \), each neighbor’s location is constrained to lie within such a probabilistic region. By aggregating the distance-based regions provided by all nodes, the leader constructs an intersection of feasible regions that captures the most likely positions of every node.
This process can be interpreted as a form of SwarmRaft's fused estimation: node \( i \)’s GNSS information defines its own anchor position, while the distance measurements extend this knowledge to constrain neighboring nodes. The leader’s role is to integrate these partial constraints from all reporting nodes, producing a consistent, network-wide probabilistic localization of the fleet.}\deleted{Peer \( j \) combines the raw GNSS reading from \( i \) with a range-based estimate to form a minimum-variance fusion. Denote the range-only estimate as}






\subsection{Fault Detection}
\replaced{In this phase, the leader evaluates the consistency of a node’s reported position by estimating its location from inter-node distances and the GNSS-based positions of its neighbors. Large discrepancies between a node’s self-reported GNSS position and the probabilistic regions derived from neighbors’ distance measurements indicate that either the reporting node’s sensors or the peers’ range measurements may be \textit{faulty}.}{In this phase, the leader evaluates each node's reported position by comparing it to a multilateration-based estimate derived from non-faulty neighbors’ reported positions and inter-node distances. A significant discrepancy indicates a potential fault in either the reporting node’s sensors or its neighbors’ measurements.}

\added{To make this precise, we define a residual, apply a statistically justified threshold, and then cast a binary vote.}

\added{Specifically, for node \( i \), the leader constructs feasible regions of its location from all non-faulty neighbor reports (as spheres/circles defined by measured distances). The reported GNSS position \( \mathbf{z}_{i,k} \) is then compared against this feasible region. We define the residual as the minimum Euclidean distance between \( \mathbf{z}_{i,k} \) and the feasible region boundary:}
\[
  e_{i,k} \;=\; \min_{\mathbf{x} \in \mathcal{R}_{i,k}} 
    \bigl\|\mathbf{x} - \mathbf{z}_{i,k}\bigr\|,
\]
\added{where \( \mathcal{R}_{i,k} \) denotes the feasible region of node \( i \)’s position at time \( k \) derived from verified neighbors.}


\added{Under nominal (non-attacked) conditions, this residual is primarily driven by sensor noise in GNSS and inter-drone ranging, so it remains small with high probability. To detect abnormal residuals, we define a threshold}
\[
  T \;=\;\mu_e \;+\; 3\,\sigma_e,
\]
\added{where \( \mu_e \) and \( \sigma_e \) are obtained via offline calibration under honest sensor operation. By Gaussian tail bounds, the probability that an honest residual exceeds \( T \) is less than 0.01. Any residual \( e_{i,k} > T \) is thus treated as strong evidence of a fault or spoofing attempt.}


\subsubsection{Local Vote Broadcast}

Based on the threshold test, \replaced{the leader assigns a binary decision for each node \( i \):}{peer \( j \) issues a binary vote}

\[
  v_{i,k}
  = 
  \begin{cases}
    +1, & e_{i,k} \le T,\\[6pt]
    -1, & e_{i,k} > T.
  \end{cases}
\]

\replaced{A decision of \( +1 \) indicates that node \( i \)’s reported position is \textit{consistent} with the probabilistic regions derived from neighbor constraints, while \( -1 \) signals a potential \textit{fault}. These decisions are then broadcast to all nodes, forming the input for the subsequent \replaced{SwarmRaft}{\textbf{SwarmRaft}} stage.}{A vote of \( +1 \) indicates \textit{honest}, while \( -1 \) signals \textit{faulty}. This local decision is then broadcast to all other peers, forming the input for the subsequent \replaced{SwarmRaft}{\textbf{SwarmRaft}} stage.}

\deleted{Leader computes votes for each node and flags their GNSS as \textit{honest} or \textit{faulty}, i.e., compute \( S_i=\sum_j v^{(j)}_{i,k} \):}


\begin{itemize}
  \item If {\( S_i \ge 0 \)}, decide $i$ is \textit{honest}.
  \item If {\( S_i < 0 \)}, decide $i$ is \textit{faulty}.
\end{itemize}

\subsection{Position Correction \& Recovery}

\added{Once a node \( i \) is flagged as \textit{faulty}, its position is recomputed using only the non-faulty neighbors' information. Let \(\mathcal{N}_i^\text{good}\) denote the set of neighbors of node \( i \) that are not flagged as faulty. Each neighbor \( j \in \mathcal{N}_i^\text{good} \) provides a reported position \(\mathbf{x}_{j,k}\) and the measured distance \(d_{i,j}\) to node \( i \).}

\added{We model the feasible location of the faulty node \( i \) as the intersection of spheres centered at each non-faulty neighbor with radius equal to the measured distance:
\[
\mathcal{S}_j = \bigl\{\mathbf{x} \in \mathbb{R}^3 \;|\; \|\mathbf{x} - \mathbf{x}_{j,k}\| = d_{i,j}\bigr\}.
\]
The intersection of these spheres defines the feasible region for the node's true position.}

\added{To compute a single position estimate \(\tilde{\mathbf{x}}_{i,k}\), we solve a nonlinear least-squares problem that minimizes the squared deviations from all spheres:
\[
\tilde{\mathbf{x}}_{i,k} = \arg\min_{\mathbf{x} \in \mathbb{R}^3} 
  \sum_{j \in \mathcal{N}_i^\text{good}} \bigl(\|\mathbf{x} - \mathbf{x}_{j,k}\| - d_{i,j}\bigr)^2.
\]
This formulation finds the point that best fits all range measurements from non-faulty neighbors, effectively performing a multilateration.}

\added{In practice, we initialize the optimization at the centroid of the non-faulty neighbors:
\[
\mathbf{x}^{(0)}_{i,k} = \frac{1}{|\mathcal{N}_i^\text{good}|} \sum_{j \in \mathcal{N}_i^\text{good}} \mathbf{x}_{j,k},
\]
and refine \(\tilde{\mathbf{x}}_{i,k}\) via iterative least-squares (e.g., using a soft-\(L_1\) loss to reduce sensitivity to residual errors). Once obtained, the node resets its INS state to \(\tilde{\mathbf{x}}_{i,k}\).}

\added{\textit{Note:} If too few non-faulty neighbors remain (e.g., due to false positives), we skip multilateration and rely solely on the node's INS propagation. This ensures that the recovery procedure remains robust even under adversarial conditions.}

\added{\textit{Note:} In rare cases, multiple honest sensors may be falsely flagged as faulty, potentially leaving too few verified neighbors for multilateration. When this occurs, we skip median/multilateration recovery and rely solely on the node’s INS propagation to update its state.}

\subsection{Security \& Fault-Tolerance Analysis}  
The security and fault-tolerance guarantees of \replaced{SwarmRaft}{\textbf{SwarmRaft}} arise from its algorithmic design, the assumptions of the threat model, and the underlying Raft consensus protocol.

\textbf{Safety:} SwarmRaft ensures safety through majority voting by enforcing the condition \( n \geq 2f + 1 \), the system guarantees that \textit{honest} nodes always outnumber any coalition of up to \( f \) \textit{faulty} or malicious nodes. This majority threshold prevents incorrect decisions from being adopted, even in the presence of adversarial behavior.

\textbf{Liveness:} In a synchronous network setting, the protocol guarantees liveness as long as a majority of nodes remain honest and responsive. The Raft algorithm enables reliable leader election and, once a leader is in place, timely progress is maintained throughout the swarm through efficient rounds of position estimation, fault detection, and recovery.

\textbf{Integrity:} All inter-node communications are authenticated using pre-distributed cryptographic keys, preventing forgery and tampering. Consensus-based validation ensures the trustworthiness of the \deleted{the} reported data. Although rare statistical anomalies can incorrectly flag an \textit{honest} sensor as \textit{faulty}, such false positives are mitigated by majority voting. In edge cases where too many nodes are flagged, the system falls back to inertial (INS) data for continuity, preserving mission integrity at the cost of reduced accuracy.

\textbf{Communication Efficiency:} SwarmRaft is designed to scale efficiently. The leader incurs a communication and computation cost of \( O(2(n-1)) \) per round, while non-leader nodes communicate only with the leader at \( O(2) \) cost. This model minimizes bandwidth usage and processing load as the swarm scales. Raft log replication further ensures consistency across all nodes.

\textbf{Fault Tolerance:} The protocol tolerates up to \( f \) Byzantine faults, including GNSS spoofing, range manipulation\added{,} and collusion. By requiring \( n \geq 2f + 1 \), SwarmRaft ensures that honest node votes dominate and that consensus decisions remain trustworthy even under adversarial conditions. The coordination of the leader of the Raft and the replicated logs reinforce\added{s} fault-tolerant behavior during ongoing operations.

\textbf{Recovery:} When faults are detected, SwarmRaft employs a robust median-based recovery strategy that replaces outlier or corrupted readings with consensus estimates derived from peer data. This allows the swarm to maintain stable formation and control even under localized sensor failures. Recovery decisions are consistently propagated via Raft’s replication mechanism, preserving swarm-wide state alignment.

\begin{figure*}[t]
    \centering
    \includegraphics[width=0.9\textwidth]{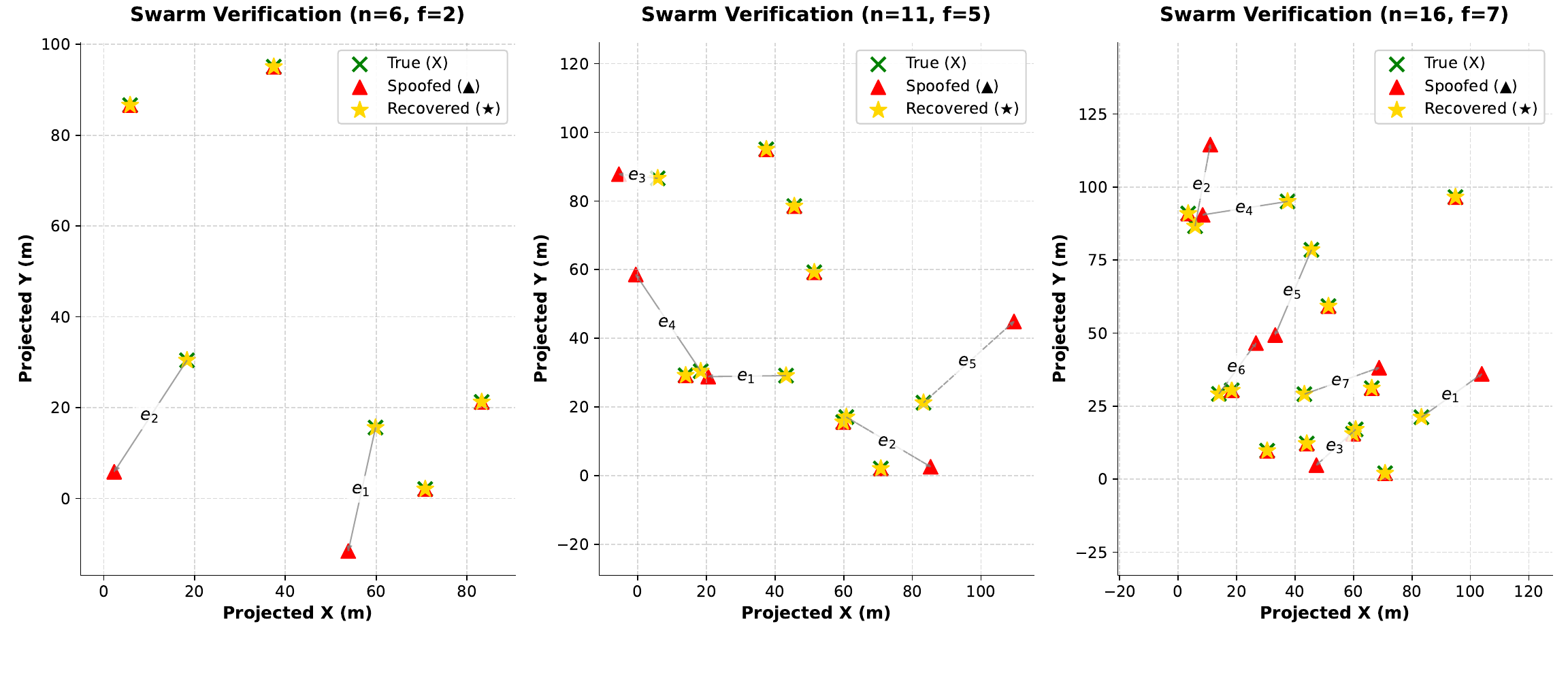}
    \caption{\added{Visualization of SwarmRaft recovery under different swarm sizes ($N = 6, 11, 16$). Yellow stars represent recovered positions using consensus and range-based fusion.}}
    \label{fig:swarmraft_viz}
\end{figure*}

\section{Prototype Implementation}

To evaluate SwarmRaft under adversarial localization scenarios, we developed a modular simulation framework in Python that models both the dynamics and vulnerabilities of drone swarms. The framework simulates each drone's ground-truth motion, inertial navigation estimates, GNSS observations (including spoofed signals), and noisy inter-drone ranging data. A dedicated leader node module performs \replaced{position verification, fault detection through voting, and recovery via iterative multilateration}{fused position estimation, fault detection through residual analysis and majority voting, and robust recovery based on peer consensus}. To support performance benchmarking, the simulator includes tools for running large-scale Monte Carlo experiments that vary swarm size and attacker configuration, producing statistical metrics including \added{mean absolute error} (MAE). Additionally, the framework supports real-time visualization and post-simulation plotting to analyze error evolution and swarm behavior.


\subsection{Algorithmic Procedure}

\added{The leader-based verification algorithm operates in two main stages. First, each drone’s reported position is cross-validated against inter-drone distance measurements. The leader tallies votes from all pairwise checks, where consistent neighbors contribute positive votes and inconsistent ones contribute negative votes. Any drone that accumulates a majority of negative votes is flagged as potentially faulty.}

\replaced{In the second stage, the leader refines the positions of flagged drones using multilateration. This involves solving a least-squares problem with respect to verified neighbors, iteratively adjusting the estimated position until convergence or until a maximum number of iterations is reached. If the corrected estimate deviates significantly from the originally reported position, the corrected value replaces the spoofed report; otherwise, the drone is reclassified as valid. This process guarantees that up to $f$ malicious or faulty drones can be corrected in a swarm of size $n \geq 2f+1$.
}{To clarify the operation of the leader-based probabilistic localization and fault detection, we summarize the steps in Algorithm~\ref{alg:neighbor-verification}.}

\begin{algorithm}[h]
\caption{\added{Leader-based Neighbor Verification with Voting and Multilateration}}
\label{alg:neighbor-verification}
\begin{algorithmic}[1]
\Require 
  Reported positions $\mathbf{x}^{\rm rep}_{i}$ from all UAVs $i = 1,\dots,n$  
  Inter-UAV distance matrix $D = [d_{ij}]$ measured from true geometry  
  Base tolerance $\epsilon > 0$, maximum iterations $k_{\max}$  
\Ensure 
  Verified positions $\mathbf{x}^{\rm ver}_i$ and fault flags for all UAVs

\State Leader collects $\mathbf{x}^{\rm rep}_{1:n}$ and distance matrix $D$

\For{each UAV $A = 1$ to $n$} \Comment{Stage 1: Voting}
    \State Initialize vote counter $v_A \gets 0$
    \For{each neighbor $B \neq A$}
        \State Compute distance from reports: $\hat{d}_{AB} = \|\mathbf{x}^{\rm rep}_A - \mathbf{x}^{\rm rep}_B\|$
        \If{$|\hat{d}_{AB} - d_{AB}| < \tau$} \Comment{consistency check}
            \State $v_A \gets v_A + 1$
        \Else
            \State $v_A \gets v_A - 1$
        \EndIf
    \EndFor
    \State Flag $A$ as faulty if $v_A < 0$
    \State Initialize $\mathbf{x}^{\rm ver}_A \gets \mathbf{x}^{\rm rep}_A$
\EndFor

\For{each UAV $A$ flagged as faulty} \Comment{Stage 2: Multilateration refinement}
    \State Select non-faulty neighbors $\mathcal{N}_A$
    \If{$|\mathcal{N}_A| < 3$}
        \State $\mathcal{N}_A \gets$ all other UAVs \Comment{fallback if insufficient anchors}
    \EndIf
    \State Initialize estimate $\mathbf{p} \gets \mathbf{x}^{\rm rep}_A$
    \For{$k = 1$ to $k_{\max}$}
        \State Update $\mathbf{p}$ by solving least-squares residuals:  
        \[
          \mathbf{p} \gets \arg\min_{\mathbf{q} \in \mathbb{R}^3} 
          \sum_{j \in \mathcal{N}_A} \big(\|\mathbf{q}-\mathbf{x}^{\rm rep}_j\| - d_{jA}\big)^2
        \]
        \If{converged} \State \textbf{break} \EndIf
    \EndFor
    \State Compute deviation $\Delta_A = \|\mathbf{x}^{\rm rep}_A - \mathbf{p}\|$
    \If{$\Delta_A > \epsilon$}
        \State $\mathbf{x}^{\rm ver}_A \gets \mathbf{p}$ \Comment{replace spoofed report}
    \Else
        \State $\mathbf{x}^{\rm ver}_A \gets \mathbf{x}^{\rm rep}_A$
        \State Mark $A$ as non-faulty
    \EndIf
\EndFor

\State \Return Verified positions $\mathbf{x}^{\rm ver}_{1:n}$ and fault flags
\end{algorithmic}
\end{algorithm}

The prototype accommodates configurable attack injection, including GNSS spoofing, range distortion, and randomized attacker behavior. The implementation serves as the experimental foundation for the visual and quantitative evaluations discussed in subsequent sections. The source code for \replaced{SwarmRaft}{SwarnRaft} is publicly available on GitHub \cite{kapeldev_swarmraft_2025}.

\section{Numerical experiments \added{and results}}\label{experiments}
\subsection{Qualitative Results for Position Recovery}

Figure~\ref{fig:swarmraft_viz} illustrates the SwarmRaft recovery mechanism across varying swarm sizes. 
\replaced{The three subplots correspond to $(n,f) \in \{(6,2), (11,5), (16,7)\}$, i.e., swarms with two, five, and seven faulty drones, respectively. 
Each visualization demonstrates that SwarmRaft is able to correctly identify spoofed drones and restore their reported positions to match ground truth via multilateration. 
For spoofed drones, the recovered position (yellow star) aligns with the true position (green cross), while the spoofed report (red triangle) is visibly displaced. 
Arrows further highlight the deviation between true and spoofed reports, making the correction visually evident.}{simulating aerial drone localization using GNSS and INS.} \added{SwarmRaft’s ability to detect and isolate faulty agents, even in small swarms with limited redundancy. The recovered positions align closely with the truth of the ground, highlighting the effectiveness of peer triangulation and consistency-based voting.} Each subfigure demonstrates how noisy or spoofed measurements are corrected through distributed voting and recovery. This visualization not only showcases concrete recovery behavior but also reveals the spatial structure of localization errors in both small- and large-scale swarms, complementing the accompanying statistical evaluation.

Each drone $D_i$ is annotated with:
\begin{itemize}
\item \textbf{Green cross (\added{$\color{green}\times$}\deleted{$\color{green}\circ$}):} 
the ground-truth position $\mathbf{x}_i^{\mathrm{true}}$.

    
\item \textbf{Red triangle (\textcolor{red}{\added{$\blacktriangle$}}):} 
\added{the spoofed reported position $\mathbf{x}_i^{\mathrm{rep}}$.}
    \deleted{\textbf{Red triangle (\textcolor{red}{$\triangle$}):}  
the INS estimate $\mathbf{x}_i^{\mathrm{INS}}$, which accumulates integration errors and noise.}

    \item \textbf{Yellow star \added{(\textcolor{yellow!50!orange}{\(\star\)}):} the recovered position $\hat{\mathbf{x}}_i$ computed via multilateration.}
    \deleted{\textbf{Yellow star (\textcolor{yellow!50!orange}{\(\star\)}):} the recovered position $\hat{\mathbf{x}}_i$ computed via the SwarmRaft algorithm.}

\end{itemize}

\replaced{The SwarmRaft protocol verifies positions through two stages. First, consistency voting checks whether reported inter-drone distances match measured distances; drones that fail majority voting are flagged as faulty. Second, flagged drones undergo multilateration using non-faulty neighbors as anchors. If the corrected estimate significantly deviates from the reported position, the spoofed report is replaced with the multilateration-based recovery. This ensures that up to $f$ compromised drones can be tolerated in a swarm of size $n \geq 2f+1$.}{The SwarmRaft protocol fuses inter-drone distance measurements and local INS readings using a consensus scheme. Each node receives fused position suggestions from its neighbors and performs majority voting to detect inconsistencies. If a node is suspected to be compromised, its position is recovered via a robust aggregation (e.g., coordinate-wise median) of neighbor estimates.}

\subsection{Quantitative Evaluation Across Swarm Sizes and Attacks}

To complement the qualitative visualizations, we conduct a large-scale statistical evaluation of SwarmRaft under varying swarm sizes and adversarial conditions. 
\replaced{For each swarm size $n \in \{3,5,7, 9,11,13,15,17\}$ and each number of attacked drones $f \in \{1,2,\dots,8\}$, we run Monte Carlo trials where $f$ drones are randomly selected as spoofed. Spoofed drones report displaced positions and manipulated ranges, while the remaining drones report correctly.}{For each swarm size $N \in \{5, 10, 15\}$ and each number of attacked drones $f \in \{1, 2, \dots, N-1\}$, we run 10 Monte Carlo trials. During each trial, $f$ drones are randomly selected to be compromised via GNSS spoofing and range manipulation. The remaining drones follow the nominal sensing model, where the GNSS and INS errors are drawn from zero-mean Gaussian distributions.}

Each trial produces two error metrics:
\begin{itemize}
  \item \textbf{Baseline Error:} 
  \replaced{The MAE of reported (potentially spoofed) positions relative to ground truth.}{The mean absolute error (MAE) calculated based on the raw GNSS readings.}
  \item \textbf{Recovered Error:} 
  \replaced{The MAE of the recovered and multilaterated positions $\hat{\mathbf{x}}_i$ produced by SwarmRaft.}{The same metrics are computed using the recovered positions \( \hat{\mathbf{x}}_i \) after consensus-based correction.}
\end{itemize}

\begin{figure}[h]
    \centering
    \includegraphics[width=0.9\linewidth]{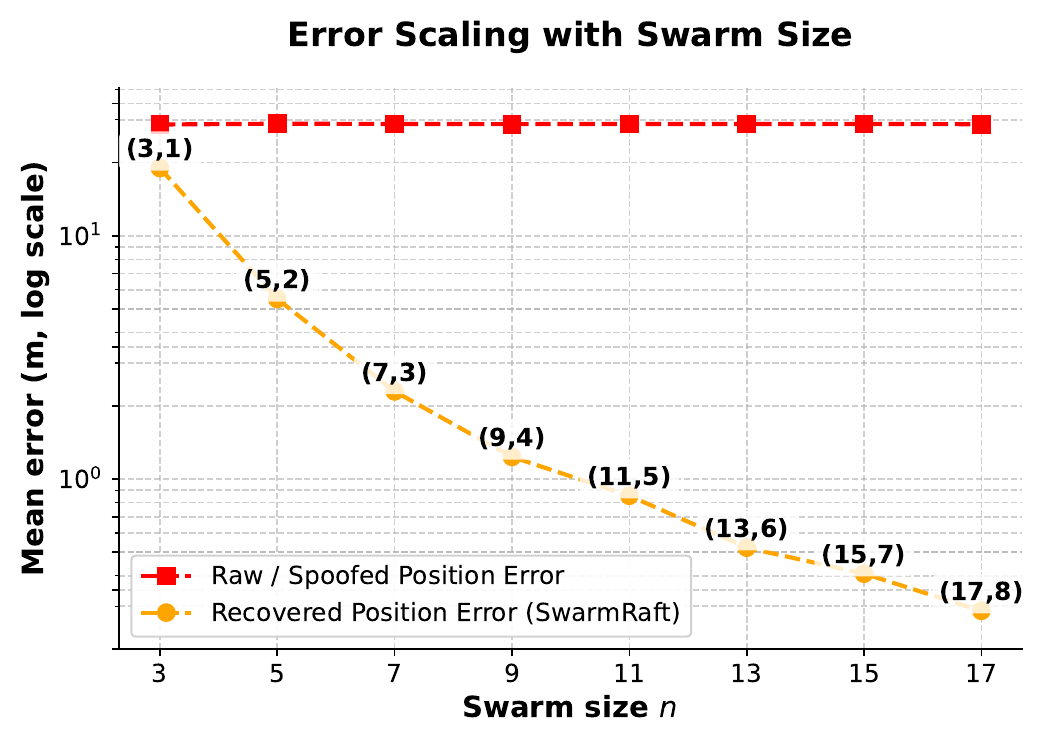}
    \caption{\replaced{Scaling performance of SwarmRaft as swarm size increases. Mean recovery error (log scale) decreases significantly with larger swarms, from ~19 m to ~0.28 m.}{Mean recovery error versus swarm size for varying numbers of faulty drones (n=2f+1), shown on a logarithmic scale and averaged over 10,000 simulations. The algorithm’s accuracy improves as swarm size increases, with errors decreasing from ~19m to ~0.28m.}}
    \label{fig:scaling_mae_rmse}
\end{figure}

\replaced{Figure~\ref{fig:scaling_mae_rmse} shows the scaling performance of SwarmRaft as swarm size grows. For a given number of faulty drones $f$, we consider the minimal swarm size $n = 2f+1$, which guarantees that the majority of drones remain non-faulty. The plot reports the mean recovery error on a logarithmic scale, averaged over $10,000$ Monte Carlo simulations. As the size of the swarm increases relative to $f$, the recovery accuracy improves dramatically, with a mean error dropping from approximately 19\,m to 0.28\,m.}{The results are aggregated in all trials and shown in Figure~\ref{fig:scaling_mae_rmse}. Each curve depicts the average error as a function of the number of attacked drones $f$, grouped by the total swarm size $N$.}

We observe that:
\begin{enumerate}
    \item \replaced{Reported (spoofed) positions quickly diverge from ground truth, reflecting the impact of adversarial manipulation.}{GNSS-only estimates degrade significantly as $f$ increases, particularly in larger swarms.}
    \item SwarmRaft consistently reduces error compared to the baseline, even when nearly half of the swarm is compromised.
    \item Recovery accuracy improves with swarm size, demonstrating scalability and resilience of the protocol.
\end{enumerate}

\begin{table*}[h]
\centering
\begin{tabular}{lccc}
\hline
\textbf{System} & \textbf{Relevant dimension} & \textbf{Complexity per update} & \textbf{Typical scale} \\
\hline
GNSS (LS solution)   & $m \approx 6$--$10$ satellites & $\mathcal{O}(m^3)$ (constant) & Few $10^2$ ops \\
INS propagation      & $n \approx 9$--$15$ states    & $\mathcal{O}(n)$             & Few $10^2$ ops \\
GNSS+INS (Kalman)    & $n \approx 15$ states         & $\mathcal{O}(n^3)$           & $\sim$3,000 ops \\
Proposed algorithm   & $N$ drones                    & Regular: $\mathcal{O}(N)$, Leader: $\mathcal{O}(N^2)$ & $200$--$6{,}000$ ops (for $N \leq 17$) \\
\hline
\end{tabular}%
\caption{Computational complexity comparison of GNSS, INS, and the proposed verification algorithm.}
\label{tab:complexity}
\end{table*}

Figure~\ref{fig:scaling_mae_rmse} further compares error distributions across trials. 
\replaced{Baseline errors show high spread, large medians, and many outliers, confirming vulnerability to spoofing. In contrast, SwarmRaft-corrected positions exhibit low medians, tight interquartile ranges, and few outliers, reflecting robustness under adversarial conditions.}{The GNSS error distribution shows a higher median, wider spread, and more extreme values, reflecting its vulnerability to spoofing and sensor faults. In contrast, the SwarmRaft-corrected estimates demonstrate a substantially reduced median error and tighter interquartile range, with fewer outliers. This indicates both improved accuracy and robustness, underscoring the protocol’s effectiveness in mitigating adversarial localization noise through consensus and peer fusion.}

\subsection{\added{Computational Complexity}}
\added{To provide a fair comparison, we summarize the computational complexity of the proposed verification algorithm against conventional GNSS and INS solutions. 
For GNSS, position computation after signal tracking reduces to solving a set of pseudorange equations, typically involving $m \approx 6$--$10$ satellites. 
This least-squares solution requires a matrix inversion of size $m \times m$, with complexity $\mathcal{O}(m^3)$, which is effectively constant in practice. 
INS propagation involves simple state integration (position, velocity, attitude), with complexity $\mathcal{O}(n)$ for a state dimension $n \approx 9$--$15$. 
When GNSS and INS are fused using a Kalman filter, the update step has complexity $\mathcal{O}(n^3)$ due to covariance updates, where $n \approx 15$ states, yielding on the order of $3{,}000$ arithmetic operations per update. }

\added{In contrast, in our proposed distributed swarm verification scheme, each regular node computes distances to all other nodes, yielding $\mathcal{O}(N)$ operations per update for swarm size $N$. 
The leader node, in the worst-case solvable scenario ($f \approx (N-1)/2$ faulty nodes), must iteratively correct $f$ faulty positions, resulting in $\mathcal{O}(fN) \approx \mathcal{O}(N^2)$.}
\added{For typical values ($N \leq 17$), this corresponds to fewer than $6{,}000$ floating-point operations, which is of the same order as a GNSS+INS Kalman update and easily handled by modern embedded processors. A summary of these computational complexities, together with the proposed SwarmRaft verification algorithm, is provided in Table \ref{tab:complexity} for ease of comparison.}

\section{Conclusion}

\added{This paper introduced SwarmRaft, a novel consensus-based protocol for resilient and decentralized UAV swarm localization in GNSS-degraded and adversarial environments. Unlike prior approaches that rely on centralized control or computationally intensive BFT protocols, SwarmRaft adapted the Raft consensus algorithm to support lightweight, crash-tolerant decision-making within resource-constrained aerial swarms. The system fused inertial navigation data with inter-drone ranging and applied distributed voting to detect and recover from sensor faults or spoofed position reports. Through extensive simulations under varying swarm sizes and attack intensities, SwarmRaft demonstrated significant improvements in both mean and median localization accuracy compared to GNSS-only baselines, particularly in scenarios with partial compromise. The protocol’s ability to maintain collective situational awareness without a central authority highlights its practical deployability for time-sensitive missions such as autonomous logistics, disaster response, and infrastructure inspection. Looking ahead, future research will focus on incorporating confidence-weighted fusion, dynamic trust scores, and asynchronous consensus strategies, as well as validating the system on physical UAV swarms in real-world settings. These advancements aim to establish SwarmRaft as a foundational architecture for fault-resilient coordination in next-generation autonomous multi-agent systems.}

\deleted{This paper presented SwarmRaft, a consensus-based protocol for resilient UAV swarm localization in GNSS-degraded and adversarial environments. By fusing inertial measurements with peer-to-peer ranging and validating positional consistency through distributed voting, SwarmRaft achieves accurate state estimation even in the presence of GNSS spoofing and range manipulation. In contrast to centralized or purely sensor-driven approaches, it integrates fault detection, consensus-based validation, and position recovery into a lightweight and scalable framework inspired by Raft.}

\deleted{Extensive simulations across varying swarm sizes and attack conditions demonstrated that SwarmRaft consistently reduces both mean and median localization errors relative to raw GNSS measurements. Notably, the protocol maintains stable performance even as the number of compromised agents increases, with larger swarms enhancing robustness through increased redundancy and more reliable peer estimates.}

\deleted{Beyond adversarial fault mitigation, SwarmRaft offers a generalizable architecture for collaborative localization under uncertainty, requiring minimal assumptions about trust, infrastructure, or communication hierarchy. Its compatibility with standard UAV sensing modalities (GNSS, INS, and ranging) and its ability to operate in fully decentralized settings support its applicability in real-world missions such as disaster response, infrastructure inspection, and autonomous logistics.}

\deleted{Future work will focus on extending the protocol with dynamic trust modeling, adaptive voting thresholds, and confidence-weighted fusion to enhance resilience against sensor drift and coordinated adversarial behavior. In addition, performance under asynchronous communication and packet loss should be evaluated, alongside validation on physical UAV swarms operating in real-world conditions. Ultimately, SwarmRaft establishes a principled foundation for robust, autonomous multi-agent systems capable of reliable operation in contested and degraded environments.}

\section*{Acknowledgment}

We acknowledge the use of ChatGPT and DeepSeek in enhancing the readability and clarity of this manuscript. These tools were employed to assist in refining language and improving the overall presentation of the content. However, the authors retain full responsibility for the integrity, accuracy, and intellectual contributions of the research at all stages.

\ifCLASSOPTIONcaptionsoff
  \newpage
\fi



\bibliographystyle{IEEEtran}
\bibliography{biblio}
%



%

\begin{IEEEbiography}
[{\includegraphics[width=1in,height=1.25in,clip,keepaspectratio]{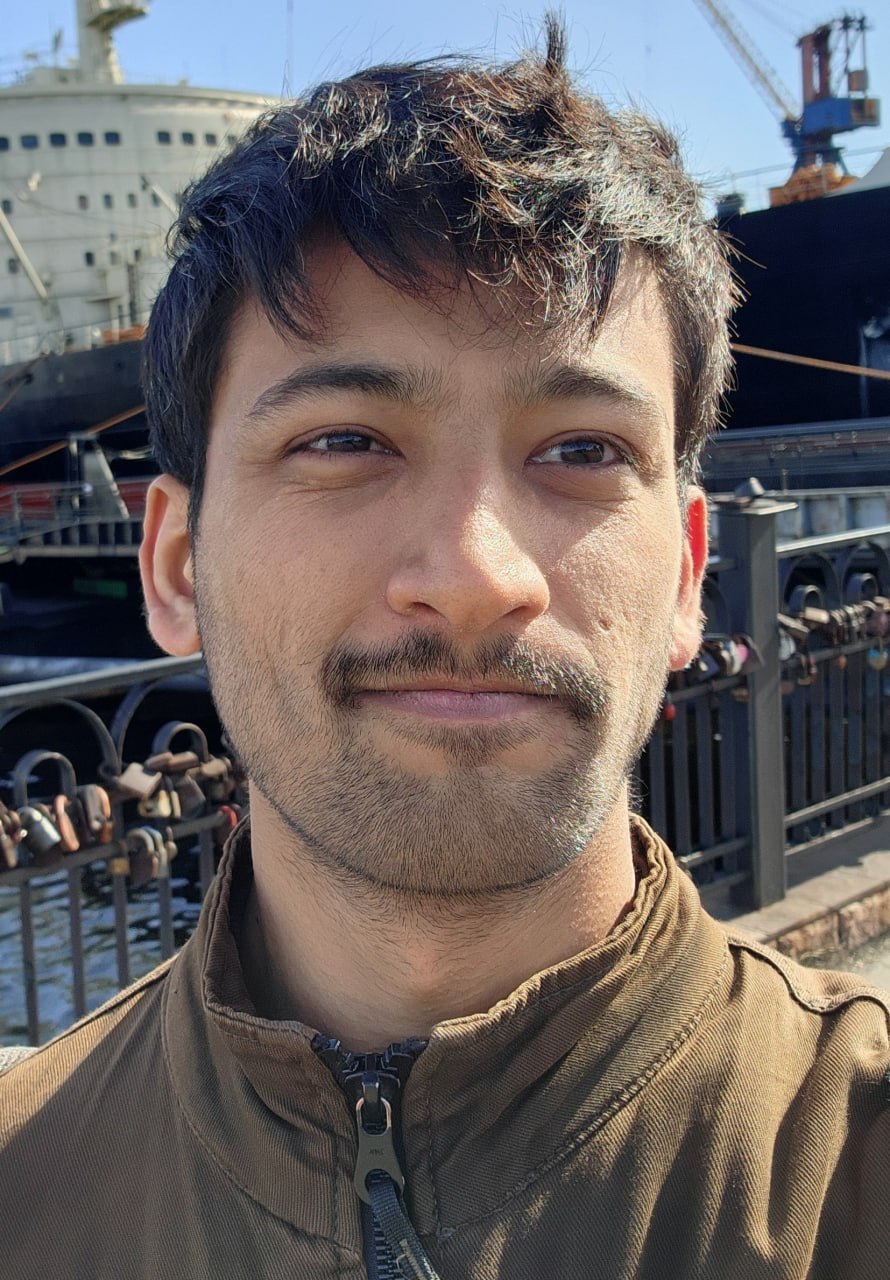}}]{Kapel Dev} is a wireless communication researcher with expertise in secure positioning, signal processing, and blockchain applications for next-generation networks. He holds an M.Sc. in Information Technology from the Skolkovo Institute of Science and Technology (Skoltech), where his thesis addressed security vulnerabilities in IEEE 802.11az ranging protocols. His work includes international research collaborations on physical-layer security, SDR-based wireless testbeds, and a co-invented U.S. patent for secure key exchange. His current research explores blockchain-integrated solutions to enhance trust and resilience in positioning systems.
\end{IEEEbiography}

\begin{IEEEbiography}
[{\includegraphics[width=1in,height=1.25in,clip,keepaspectratio]{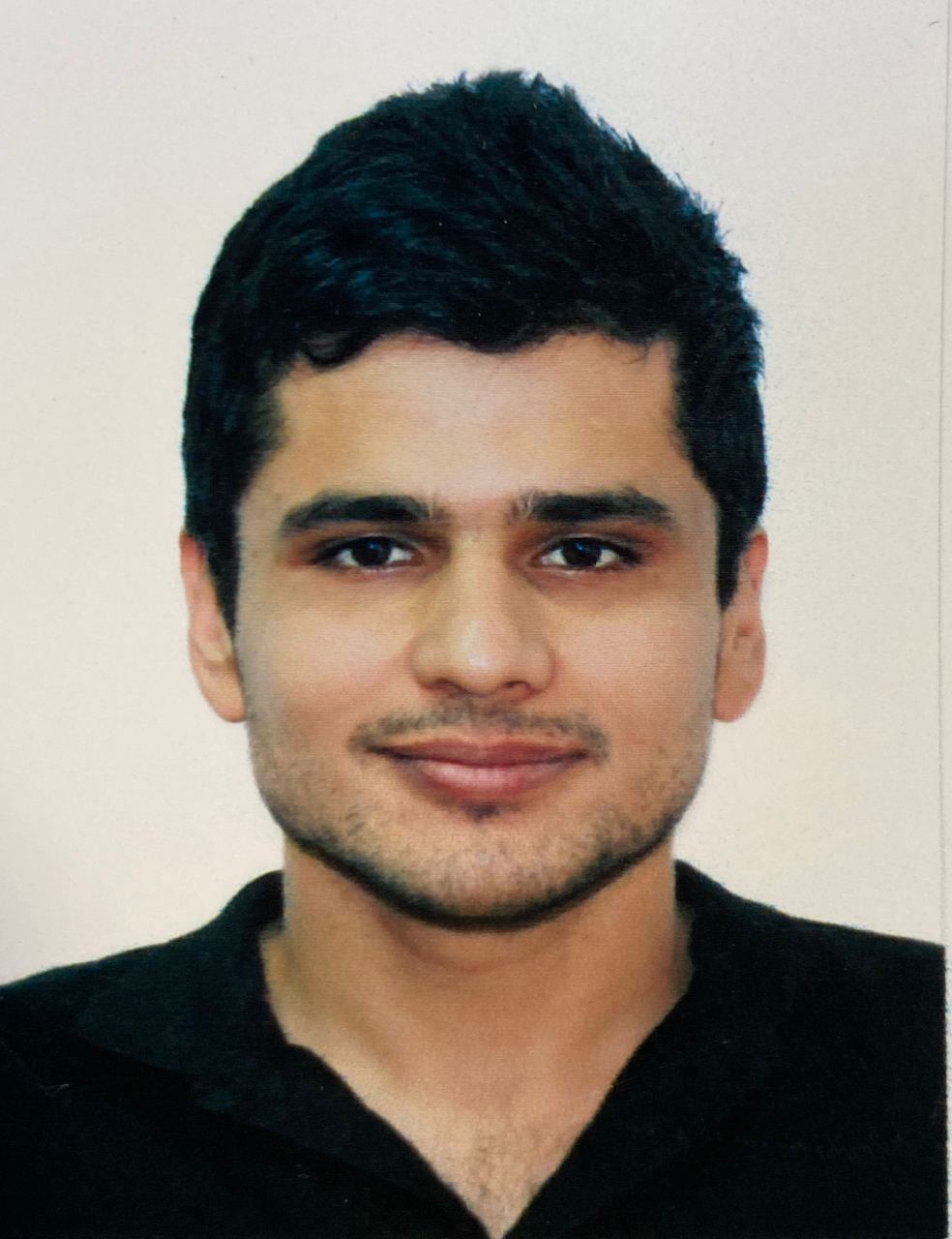}}]{Yash Madhwal}
(Member, IEEE) is a Research Scientist at the Skolkovo Institute of Science and Technology (Skoltech), where he specializes in applying blockchain technology to solve supply chain challenges. He earned his Ph.D. in 2024, with a focus on blockchain applications in supply chain systems. Yash has authored several scientific papers in which he developed prototypes of blockchain-based decentralized applications (DApps) designed to address real-world industrial problems. He serves as a teaching assistant for the course "Introduction to Blockchain" and regularly conducts technical seminars that demonstrate practical methods for building blockchain applications. Additionally, he serves as a guest lecturer at various universities, delivering introductory lectures on blockchain technology and its potential applications.
\end{IEEEbiography}

\begin{IEEEbiography}
[{\includegraphics[width=1in,height=1.25in,clip,keepaspectratio]{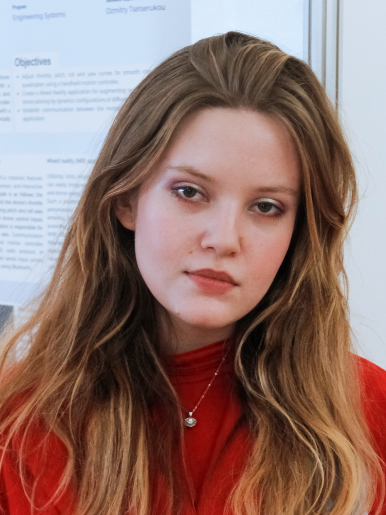}}]{Sofia Shevelo} is a M.Sc. graduate in Engineering Systems from the Skolkovo Institute of Science and Technology (Skoltech), Moscow, Russia. Her research focuses on human–robot interaction (HRI), haptics, robotics, machine learning, and computer vision. Her research explores gesture-based interfaces to improve the intuitiveness and user experience of drone operating systems. She collaborates with the Artificial Intelligence in Dynamic Actions team at Skoltech and contributes to projects involving multi-modal user interfaces for robotics platforms.
\end{IEEEbiography}

\begin{IEEEbiography}
[{\includegraphics[width=1in,height=1.25in,clip,keepaspectratio]{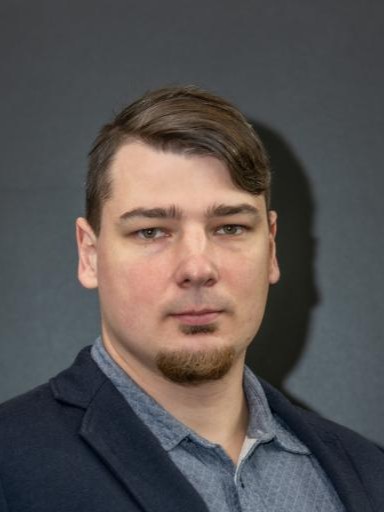}}]{Pavel Osinenko} is an Associate Professor at the Skolkovo Institute of Science and Technology (Skoltech), where he leads the AIDA (AI in Dynamic Action) research group. He earned his Doctor habilitatus degree in 2025 from Technische Universität Chemnitz, Germany, focusing on reinforcement learning (RL) with formal guarantees for safety, stability, and robustness. His research interests center on the intersection of RL and control theory, particularly the development of trustworthy AI for dynamical systems. He has authored numerous peer-reviewed publications and proposed several RL algorithms, including the CALF (Critic as Lyapunov Function) framework. At Skoltech, he teaches graduate courses and supervises research in machine learning, RL, and control systems. He also conducts technical seminars and contributes to interdisciplinary projects aimed at building reliable, provable, and safe AI systems for real-world applications.
\end{IEEEbiography}

\begin{IEEEbiography}[{\includegraphics[width=1in,height=1.25in,clip,keepaspectratio]{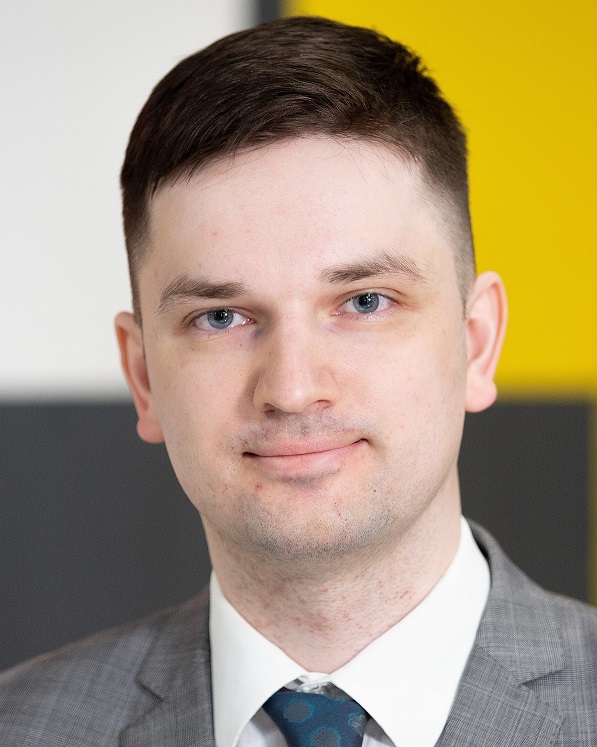}}]{Yury Yanovich} (Member, IEEE) received his Bachelor's (Honours) and Master's (Honours) degrees in Applied Physics and Mathematics from the Moscow Institute of Physics and Technology, Moscow, Russia, in 2010 and 2012, respectively. He received his Ph.D. in Probability Theory and Mathematical Statistics from the Institute for Information Transmission Problems, Moscow, Russia, in 2017. He is currently an Assistant Professor at the Skolkovo Institute of Science and Technology, Moscow, Russia. Yury is the author of the Exonum consensus protocol and has been a lecturer in the Introduction to Blockchain course at top Russian universities since 2017. His research interests include blockchain, consensus protocols, privacy, and their applications.
\end{IEEEbiography}

\end{document}